\documentclass[final,1p,times]{elsarticle}

\usepackage{amssymb}
\usepackage{amsmath}
\usepackage{amsthm}
\usepackage{color}
\usepackage{ulem} 
\usepackage{soul} 

\usepackage{lineno}
\usepackage{graphicx}
\usepackage{subcaption}
\journal{J. Quant. Spectrosc. Radiat. Transfer}

\begin{document}

\begin{frontmatter}



\title{Comb-locked cavity ring-down spectroscopy of CO$_2$
at 2-$\mu$m wavelength}


\author{Muhammad Asad Khan} 
\author{Vittorio D'Agostino} 
\author{Stefania Gravina} 
\author{Livio Gianfrani} 
\author{Antonio Castrillo\corref{cor1}} 
\cortext[cor1]{Corresponding author}
\ead{antonio.castrillo@unicampania.it}

\affiliation{organization={Dipartimento di Matemtatica e Fisica, Università degli Studi della Campania "Luigi Vanvitelli"},
            addressline={Viale Lincoln, 5}, 
            city={Caserta},
            postcode={81100}, 
            country={Italy}}

\begin{abstract}
We report on a comb-locked cavity ring-down spectrometer developed for high-precision, SI-traceable, molecular spectroscopy of air-broadened CO$_2$ gas samples. The experimental setup relies on the use of a singly-resonant optical parametric oscillator that acts as an intermediate link between a 2 $\mu$m external-cavity diode laser and an optical frequency comb stabilized against a GPS-disciplined Rb-clock. Absorption spectra of the R(50) ro-vibrational component of the CO$_2$ 20012-00001 band have been recorded with high precision and fidelity. As a result of a refined spectral analysis, based on the implementation of the modified Hartmann-Tran profile, line center frequencies, pressure broadening and pressure shifting coefficients have been determined. Finally, we demonstrate the measurement of CO$_2$ mole fractions with a subpromille statistical uncertainty.
\end{abstract}







\end{frontmatter}


\section{Introduction}
\label{intro}
High-quality molecular spectroscopic parameters (such as line positions, intensities, and pressure-broadening coefficients) are essential for the quantitative analysis of the interaction between radiation and matter across a variety of scientific and industrial applications.  In recent years, the combination of cavity-enhanced spectroscopic methods with optical frequency-comb technologies has allowed to add metrology-grade qualities to spectroscopic determinations in the gas phase \cite{Gianfrani2024}. As a result, absolute transition frequencies of gaseous molecules can be measured with relative uncertainties as low as a few parts over 10$^{11}$, even approaching the 10$^{-12}$ level in some cases \cite{Reed2020, Aiello2022, Castrillo2023}. Similarly, extraordinary progress has been made in our ability to perform line intensity determinations, which are notoriously much more complicated \cite{Polyansky2015, Odintsova2017, Odintsova2020}. Recently, subpromille measurements of CO overtone line intensities have been demonstrated using absorption- and dispersion-based cavity-enhanced techniques \cite{Bielska2022}. Such high accuracy is particularly useful for atmospheric monitoring applications, including satellite observations of the various constituents of the Earth’s atmosphere. More particularly, continuous monitoring of atmospheric carbon dioxide remains a critical priority for climate science, as global concentrations have recently approached the level of 430 parts per million (ppm). Satellite observatories, such as those based on the NASA OCO-2 and OCO-3 instruments (OCO standing for Orbiting Carbon Observatory) provide column-averaged CO$_2$ densities with a precision of about 1 ppm \cite{Eldering2019}. OCO-3 operates at three near-infrared wavelengths to probe the O$_2$ A-band at 0.76 $\mu$m and the so-called weak CO$_2$ and strong CO$_2$ bands in the vicinity of 1.61 and 2.06 $\mu$m, respectively. The accuracy of the retrieved data obviously depends on the uncertainty of the CO$_2$ line intensities.

Highly accurate line intensities are also important for temperature metrology by means of line intensity ratio thermometry (LRT) \cite{Santamaria2019}. In this respect, an interesting development has recently been reported by Lisak and co-workers \cite{Lisak2025}, who demonstrated thermodynamic temperature determinations with a global uncertainty of 82 ppm (24 mK at 296 K) using CO (3-0) vibrational band lines and cavity mode dispersion spectroscopy.
In this paper, we report on a comb-locked cavity ring-down spectrometer expressly designed to produce metrology-grade spectroscopic parameters for the spectral features of carbon dioxide in the 2-$\mu$m wavelength region.

\section{Experimental setup}
\label{expset}
\begin{figure}
\centering
\includegraphics[width=1\textwidth]{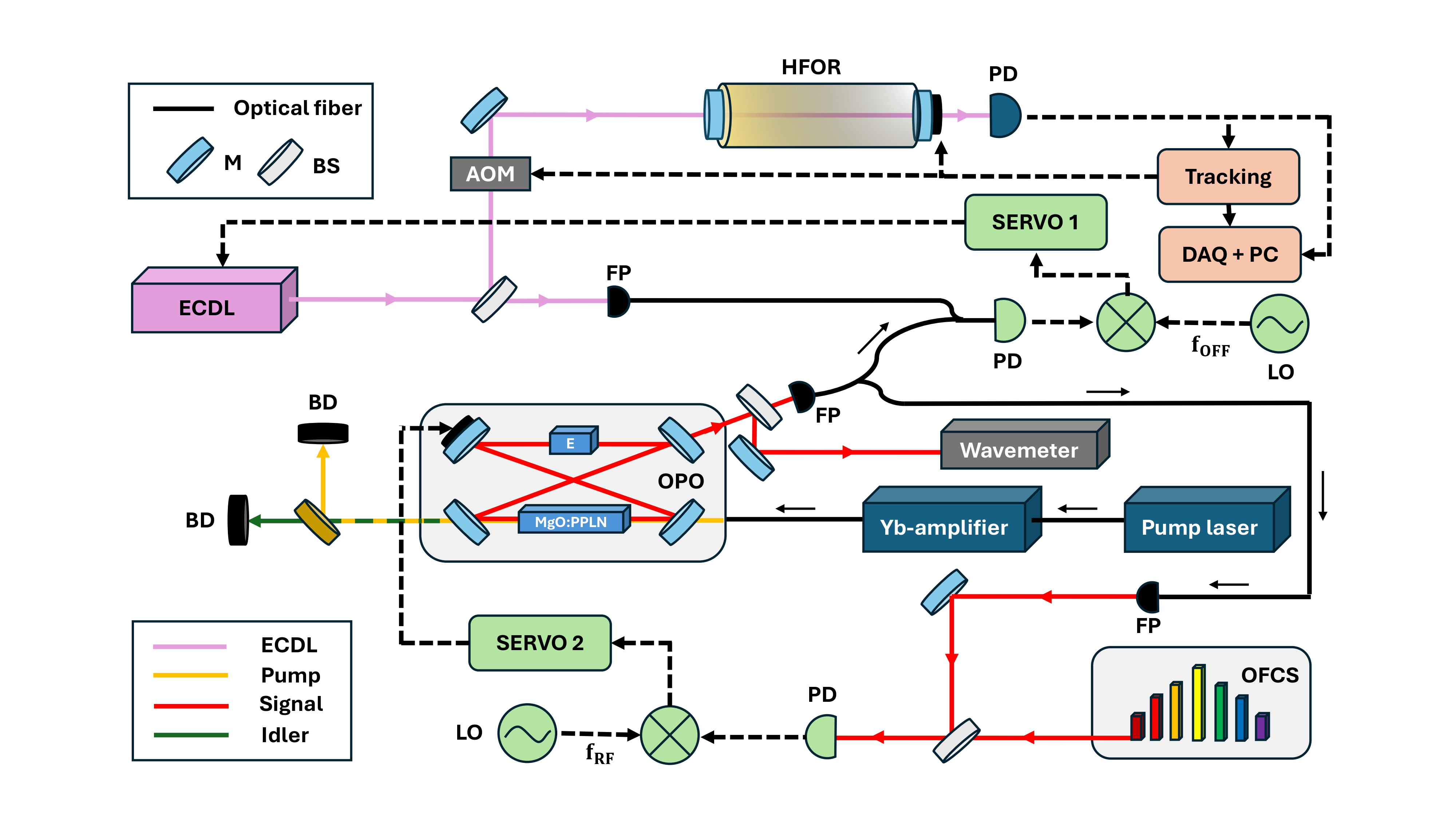}
\caption{Schematic of the experimental setup. OFCS stands for optical frequency comb; ECDL, external-cavity diode laser; OPO, optical parametric oscillator; E, etalon; PD, photodiode; AOM, acoustic-optic modulator; BD, laser beam dump; DAQ,
acquisition board; HFOR, high-finesse optical resonator; FP, fiber port; PC, personal computer; M, mirror; BS, beam-splitter; LO, local oscillator. Black dashed arrows indicate electrical connections, while continuous black lines stand for optical fiber.}
\label{fig1}
\end{figure}

Absorption spectra were acquired using a comb-locked cavity ring-down (CL-CRDS) spectrometer, whose first implementation is discussed in Ref. \cite{Dagostino2025}. The experimental setup, depicted in Fig. \ref{fig1}, consists of four modules: a 2-$\mu$m widely tunable external-cavity diode laser (ECDL) used to probe the gas sample, a GPS-disciplined optical frequency comb synthesizer (OFCS),  a singly-resonant continuous-wave comb-locked optical parametric oscillator (OPO), which is used as an intermediate optical link between the OFCS and the ECDL, and a high-finesse optical resonator (HFOR) that continuously tracks the ECDL when this latter is scanned across a selected transition.

The ECDL emits 20 mW on a single mode in the spectral range between 1980 and 2100 nm, with a full-width at half-maximum (FWHM) of approximately 200 kHz at an integration time of 1 ms. The laser beam is split into two parts. One portion is sent to a fiber-coupled 12.5-GHz bandwidth InGaAs photodiode for the detection of the beat note with the signal output of the OPO source. The remaining portion of the ECDL radiation, after passing through an acousto-optic modulator (AOM), is coupled into the HFOR, with its frequency down-shifted by f$_{AOM}$ = 80 MHz. The light transmitted by the HFOR is monitored by a 1-MHz InGaAs photodetector. At a fixed ECDL frequency, a tracking servo-loop circuit actuates the piezoelectric transducer (PZT) mounted on the output mirror of the HFOR, thus scanning the cavity length until a resonance condition is achieved \cite{Fasci2018}. When the intracavity power reaches a predefined threshold level, a TTL trigger signal is generated and sent to the AOM to abruptly switch off the laser light. This is accomplished by means of a RF switch (RF BAY, model RFS-1) ensuring an extinction ratio higher than 45 dB. The TTL triggers an exponential decay of the intracavity field, allowing a ring-down event to be recorded. The ring-down signals are acquired using a digital acquisition board (DAQ) with a 16-bit vertical resolution and a sampling rate of 5 MS/s.

The OPO is pumped by a distributed-feedback diode laser emitting at 1064 nm and amplified by a Yb-doped fiber amplifier to an output power of up to $\sim$10 W. The OPO bow-tie ring cavity is formed by four mirrors, which are highly reflective ($\sim$99.9\%) at the signal wavelength (1400-2070 nm) and transmit the pump and idler wavelengths. The pump radiation is focused into a nonlinear MgO:PPLN crystal, which is placed inside the OPO cavity. The crystal temperature is actively stabilized, and the desired phase-matching condition is obtained by selecting the appropriate poling period. To reduce the probability of mode-hops, a solid etalon is placed in the secondary focus of the OPO cavity. Coarse tuning of the wavelength is done by selecting the appropriate poling period, while fine tuning and scanning of the signal frequency are performed by actuating one of the cavity mirrors mounted on a PZT. The signal frequency is directly influenced by the crystal temperature and position, the etalon angle, and the cavity length, whereas the idler frequency is determined by energy conservation and is equal to the difference between the pump and signal frequencies. The output powers of the signal and idler are approximately 3 and 1 W, respectively. For the purposes of the present work, the idler radiation is not used, while the signal is employed as the reference laser, exhibiting a full-width at half-maximum of about 230 kHz \cite{Dagostino2025}. Once the desired frequency is reached, the signal beam is stabilized against a comb tooth with an offset frequency of f$_{OFF}$ = 20 MHz. The resulting correction signal actively controls the OPO cavity length via the PZT, with a bandwidth of a few tens of hertz.

The self-referenced OFCS provides a supercontinuum spanning in the 1050-2100 nm spectral range and operates with a repetition-rate frequency $f_{REP}$=250 MHz, while its carrier-envelope offset frequency is stabilized at $f_{CEO}$=20 MHz. Both frequencies are phase-locked to a GPS-disciplined Rb clock. The in-loop relative stability of the OPO signal frequency with respect to the OFCS is of the order of 10$^{-12}$ at 1 s of integration time \cite{Dagostino2025}. 

The locking of the OPO signal to the optical frequency comb has a twofold purpose: first, it minimizes potential long-term frequency drifts, and second, it enables the determination of its absolute frequency, $f_{OPO}$. In fact, the OPO signal frequency can be determined using the following equation: $f_{OPO} = N \times f_{REP} \pm f_{CEO} \pm f_{OFF}$, where $N$ denotes the order of the selected comb tooth, which can be easily determined using a wavemeter with an accuracy of 30 MHz.

The ECDL frequency, $f_{ECDL}$, can be tuned across the selected transition by adopting the strategy described in Ref. \cite{Castrillo2010}. Specifically, the beat-note frequency between the ECDL and the OPO signal is compared with that of a local radio-frequency oscillator, $f_{RF}$, allowing the frequency offset between the two lasers to be actively controlled by a dedicated servo loop. Consequently, stepwise variations of $f_{RF}$ result in a highly accurate scan of the ECDL frequency, spanning up to 10 GHz. Under these conditions, the absolute frequency of the probe laser is given by $f_{ECDL} - f_{AOM}$, i.e., $N \times f_{REP} \pm f_{CEO} \pm f_{OFF} \pm f_{RF} - f_{AOM}$. All the frequency terms appearing in the latter equation were referenced to the GPS-disciplined time base during the experiment. It is worth noting that the signs of $f_{OFF}$, $f_{CEO}$, and $f_{RF}$ could be positive or negative, depending on the experimental conditions. In particular, the correct signs of $f_{CEO}$ and $f_{OFF}$ could be readily determined by slightly varying $f_{REP}$ and $f_{CEO}$ and monitoring the resulting variation of $f_{OFF}$, while the sign ambiguity on $f_{RF}$ could be removed by comparing the probe laser wavelength with that of the reference laser.

The HFOR consists of of two plano–concave high- reflectivity mirrors separated by a Zerodur spacer with a length of 43 cm, corresponding to a free spectral range of about 350 MHz. The cavity is equipped with a 13.3 kPa absolute pressure gauge and a calibrated pt-100 platinum-resistance thermometer, enabling accurate measurements of the gas pressure and cavity temperature, respectively. Both sensors have an accuracy of 0.05\%. The two mirrors have a radius of curvature of 1 m and a nominal reflectivity exceeding 99.99\%. Under empty-cavity conditions, the corresponding ring-down time, $\tau_0$, is about 18 $\mu$s. It is important to point out that, compared to the results obtained in Ref. \cite{Dagostino2025}, a decreases of the ring-down time has been observed, most probably due to a noticeable dependence of the mirrors’ reflectivity on the laser wavelength. In fact, in Ref. \cite{Dagostino2025} the spectrometer operated at 4999 cm$^{-1}$, while in this work the spectral region around 5007 cm$^{-1}$ was investigated. From the optical power at the output of the cavity, in correspondence of the trigger threshold, we estimate that the intracavity intensity is several orders of magnitude less than the saturation intensity predicted for the transition studied here. Therefore, saturation effects were completely negligible.

The entire experiment is controlled by a LabVIEW-based software. In particular, absorption spectra are acquired by scanning the ECDL frequency over a range of approximately 3 GHz. At each frequency step, 20 consecutive ring-down events are recorded and the corresponding decay times are averaged. As a result, a single absorption spectrum is acquired in less than 40 s. Once a complete laser scan has been performed and the ring-down signals have been processed, the measured decay times are converted into frequency-dependent cavity losses, $\frac{1}{c\tau(\tilde{\nu})}$, being $c$ the speed of light in vacuum (expressed in cm/s).

\section{Spectral analysis}
\label{specanalysis}
Once an absorption spectrum is recorded, it is analyzed through a nonlinear least-squares fit to the following
function:
\begin{equation}
\label{abs_loss}
\frac{1}{c\tau(\tilde{\nu})} =  \frac{1}{c\tau_0}+P_1\times(\tilde{\nu}-\tilde{\nu}_0) + \alpha_{TOT}\times g(\tilde{\nu}-\tilde{\nu}_0),  
\end{equation}
$\frac{1}{c\tau_0}$ being a term that represents the losses from the empty cavity and absorption from the wings of all non-resonant lines, $P_1$ a parameter that takes into account possible variations due to a slightly linear dependence of the mirrors’ reflectivity on the laser frequency, $\alpha_{TOT}$ the integrated absorption coefficient (in cm$^{-2}$), and g($\tilde{\nu}$-$\tilde{\nu}_0$) the normalized lineshape function (in cm) centered at $\tilde{\nu}_0$ (in cm$^{-1}$).

Spectral analysis was performed by using the so-called modified Hartmann-Tran (mHTP) profile whose physical basis and mathematical form is extensively described in Ref. \cite{mHTP}. Here, it useful to remind that the mHTP profile involves six collisional  parameters, namely the speed-averaged pressure broadening, $\Gamma_0$, and shift, $\Delta_0$, the corresponding speed dependence parameters ($\Gamma_2$ and $\Delta_2$), the real, $\nu_{opt}^r$, and imaginary, $\nu_{opt}^i$, parts of the Dicke narrowing parameter. Additionally, the mHT profile involves the Doppler-broadening parameter, $\Gamma_D=\frac{\tilde{\nu}_0}{c}\sqrt{2\ln{2}\frac{k_BT}{m}}$, where $k_B$ represents the Boltzmann constant, $T$ the gas temperature, and $m$ the molecular mass. $\Gamma_0$ and $\Gamma_D$ are the half width at half maximum (HWHM) of the Lorentz and Doppler broadening components, respectively.

For the aim of this work, absorption spectra, acquired at nine different pressures (from 667 to 4000 Pa), are analyzed independently, keeping fixed the Doppler width to the expected value on the basis of the measured temperature, and remaining as variable parameters $\Gamma_0$ and $\nu_{opt}^r$. Pressure-shifting parameters and $\nu_{opt}^i$ are set to zero, while, according to the approach proposed by Fleurbaey \textit{et al.} \cite{FLEURBAEY2020}, $\Gamma_2$ was kept fixed (at each pressure) to the value that makes equal to 0.073 the ratio $a_w=\frac{\Gamma_2}{\Gamma_0}$. It must be noted that $a_w$ was calculated according to Ref. \cite{LISAK2015}, taking into account that ambient air, namely the collisional partner of CO$_2$ in our experiment, can be considered a gas mixture roughly containing a $\zeta_{N_2}$=0.8 fraction of N$_2$, while the renaming gas is O$_2$. More specifically, it comes that:
\begin{equation}
a_w= (1-n)\frac{2}{3}\Bigg[\zeta_{N_2} \frac{m_{N_2}}{1+m_{N_2}/m_{CO_2}}+(1-\zeta_{N_2})\frac{m_{O_2}}{1+m_{O_2}/m_{CO_2}}\Bigg],    
\end{equation}
where $n$=0.74 is the temperature exponent of the collisional broadening coefficient and $m_{N_2}$, $m_{O_2}$, and $m_{CO_2}$, are the N$_2$, O$_2$, and CO$_2$ masses, respectively. It should be noted that the $a_w$ value adopted in the fits agrees with those determined by D. Mondelain, \textit{et al.} in Ref. \cite{MONDELAIN2023} in similar experimental conditions. The other parameters in Eq. \ref{abs_loss}, namely $\tilde{\nu}_0$, $P_1$, $\alpha_{TOT}$, and $\tau_0$, are also treated as free.

It is useful to recall that with such a choice, the mHT profile becomes a quadratic speed-dependent Nelkin-Ghatak profile, circumstance that occurred also in other studies concerning the N$_2$-, air- and self- broadened near-infrared spectrum of CO$_2$ \cite{FLEURBAEY2020,MONDELAIN2023,Odintsova2017}. 

Spectral analysis was performed by means of a nonlinear least-squares fit under the MATLAB environment using the codes provided in Ref. \cite{mHTP}. For the sake of clarity, we highlight that the choice of including the Dicke narrowing parameter in the fitting procedure makes the $\beta$-correction included in the mHT profile effective.

Note that in the fitting procedure, absorption due to the presence of a weak water line (at 5007.704330 cm$^{-1}$), in the neighborhood of the targeted CO$_2$ transition, was taken into account. In particular, it was modeled as an interfering line using a Voigt profile and spectroscopic data derived from Ref. \cite{HITRAN}. Ignoring the contribution due to water absorption slightly deteriorates the quality of the fits, while leaves unchanged the retrieval of relevant spectroscopic parameters.

\section{Results and discussion}
\label{results}
\begin{figure}
\centering
\includegraphics[width=0.8\textwidth]{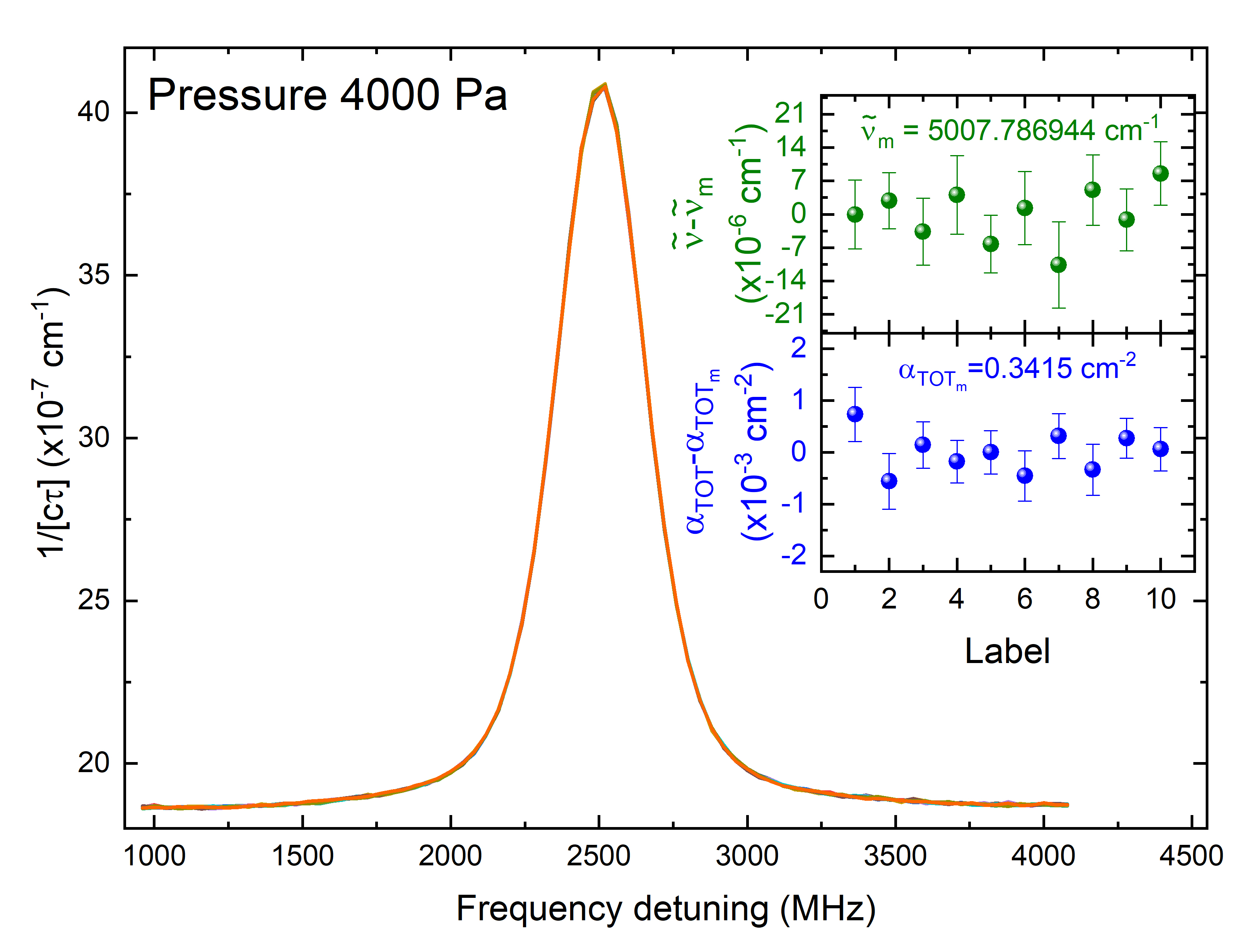}
\caption{CO$_2$ absorption spectra resulting from 10 repeated acquisitions, in coincidence with the R(50) line of the 20012-00001 band (HITRAN notation) at 5007.787078 cm$^{-1}$. The upper inset on the right quantifies the dispersion of the retrieved line-center frequencies. A 1-$\sigma$ fluctuation of 180 kHz was found. As for the vertical scale, the stability is better than 0.1\%, without signal averaging, as shown in the lower inset on the right.}
\label{fig2}
\end{figure}

In order to test the overall stability of the spectrometer, we recorded the absorption spectra resulting from ten consecutive scans of the laser frequency across the R(50) CO$_2$ absorption line, the gas samples being extracted from the laboratory air. This was done over the entire investigated pressure range. In particular, Fig.\ref{fig2} reports overlapped spectra at a pressure of 4000 Pa and at a temperature of (299.59$\pm$0.06) K. Here, each laser scan was about 3.2-GHz wide as obtained with 81 steps of 40 MHz each. So doing, we could quantify the frequency stability of our spectrometer, over a time span of about 4 min. A 1-$\sigma$ fluctuation of 180 kHz was found for the line-center frequency, corresponding to a relative stability of 1.2$\times$10$^{-9}$. Most probably, this is partially limited by the signal-to-noise ratio (SNR) of each individual spectrum, and by the resolution of the frequency axis. However, during the time in which these spectra were acquired, no frequency drift was evidenced, demonstrating the rather good frequency stability of our spectrometer. A similar behavior was observed also in the case of the integrated absorption coefficient, as shown in the lower inset of Fig. \ref{fig2}. These outcomes demonstrate that the requirements for spectra averaging over ten consecutive scans were satisfied.

\begin{figure}
\centering
\includegraphics[width=0.8\textwidth]{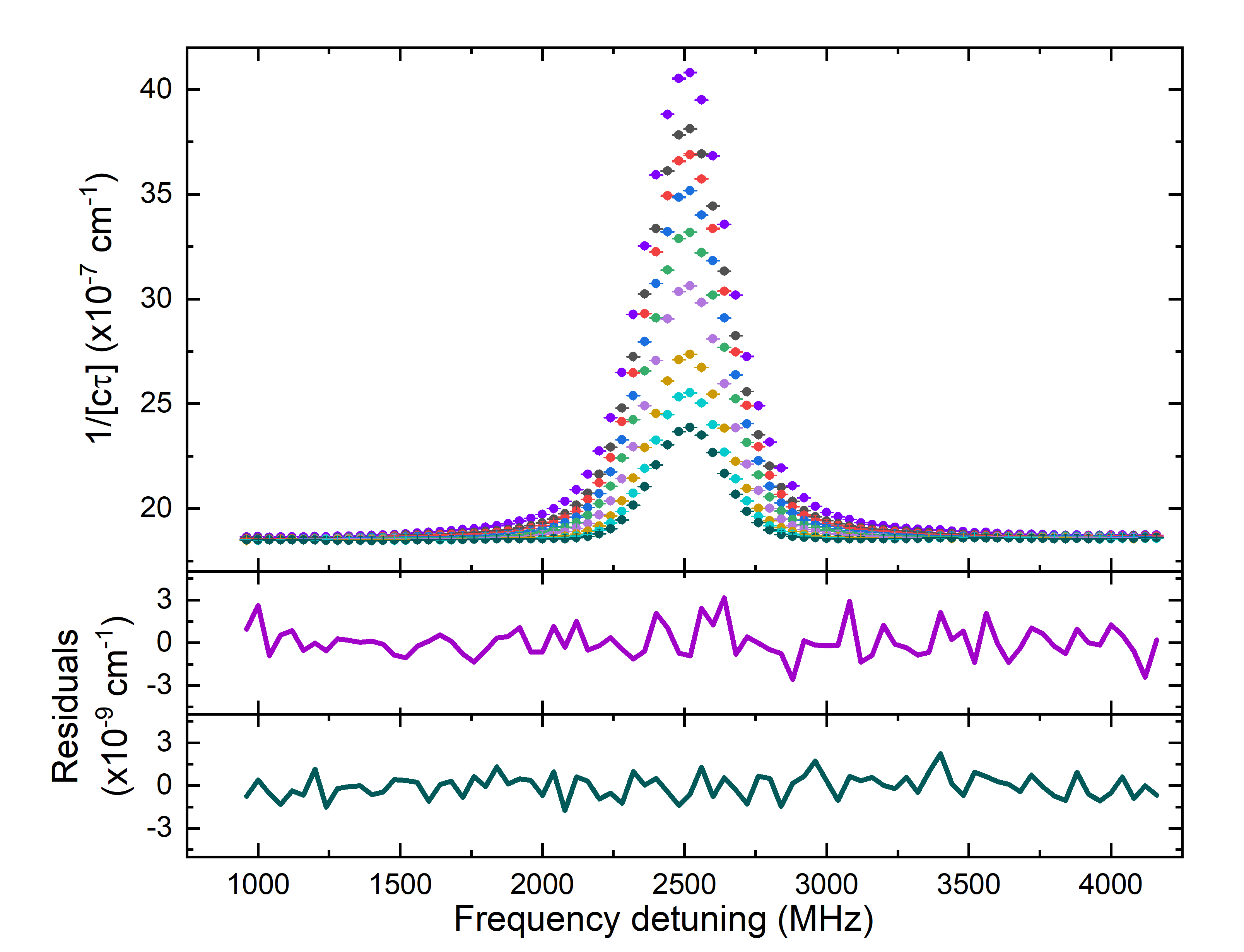}
\caption{Upper panel: Averaged CO$_2$ absorption spectra recorded at 9 different pressures of ambient air, from 667 to 4000 Pa. For each spectrum, the observed quality factors ranged from 2750 to 3600. (Quality factor is defined as the ratio of the peak absorption to the rms of the fit residuals). Middle panel: fit residuals at the highest
pressure. Lower panel: residuals at the lowest pressure.}\label{fig3}
\end{figure}

The upper panel of Fig. \ref{fig3} shows the absorption spectra resulting from averaging over ten consecutive acquisitions at each of the nine different pressure values. These spectra were interpolated by using Eq. \ref{abs_loss}, according to the procedure described in Section \ref{specanalysis}. Fig. \ref{fig3} also shows, in the middle and lower panels, fit residuals for the highest and lowest pressures, respectively. The root-mean-square (rms) values are 1.1 and 0.8$\times$10$^{-9}$ cm$^{-1}$, respectively, limited only by the experimental noise. Similar results were obtained over the entire range of pressures. 

\begin{figure}
    \begin{subfigure}{\textwidth}
    \centering
    \includegraphics[width=0.8\textwidth]
    {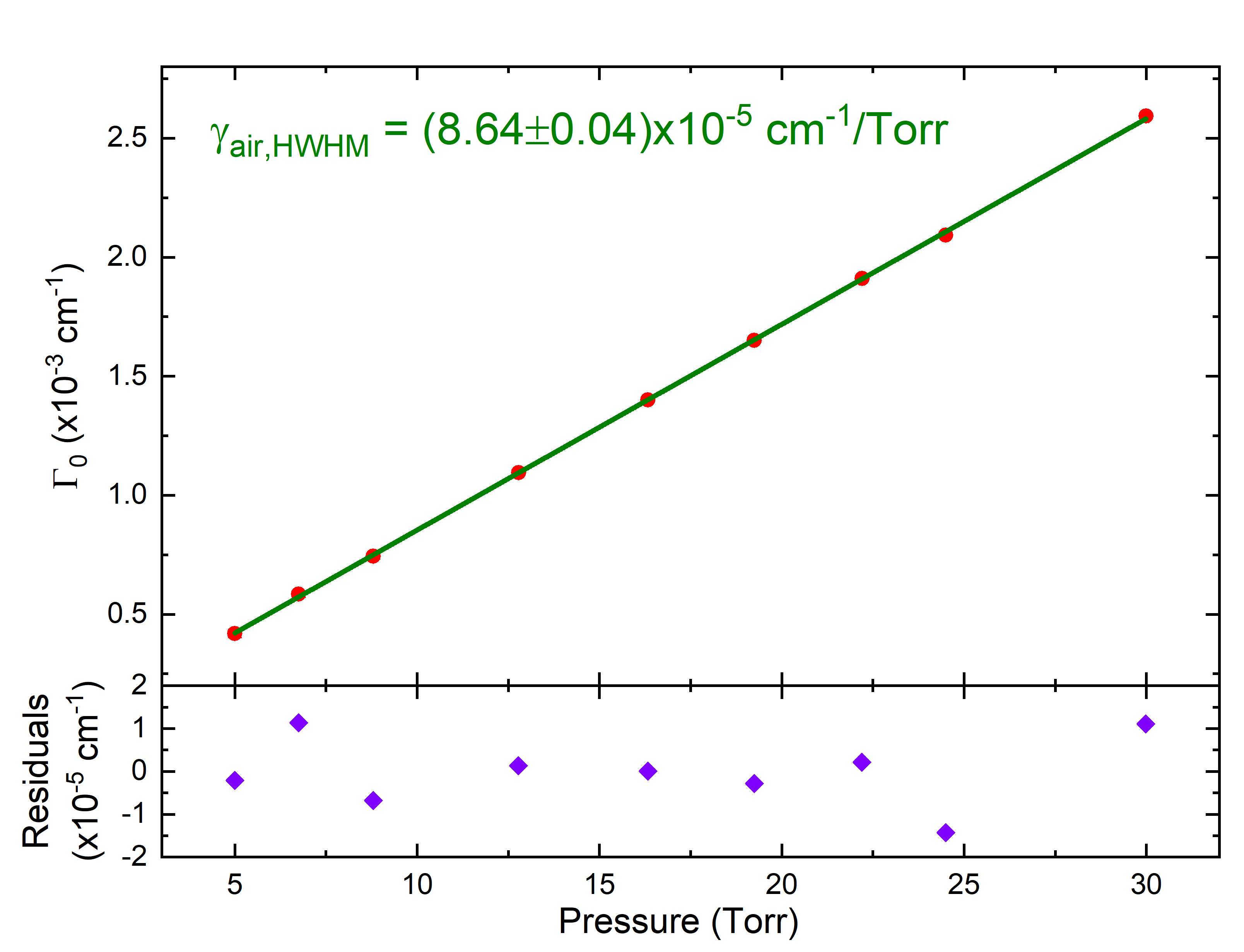}
    \caption{Upper panel: HWHM Lorentzian component as a function of the gas pressure. The slope of the best-fit line provides the air-broadening coefficient expressed in cm$^{-1}$/Torr. Lower panel: residuals from the linear fit, being their rms value 8.1$\times$10$^{-6}$ cm$^{-1}$.}
    \label{fig4a}
    \end{subfigure}
    \begin{subfigure}{\textwidth}
    \centering
    \includegraphics[width=0.8\textwidth]{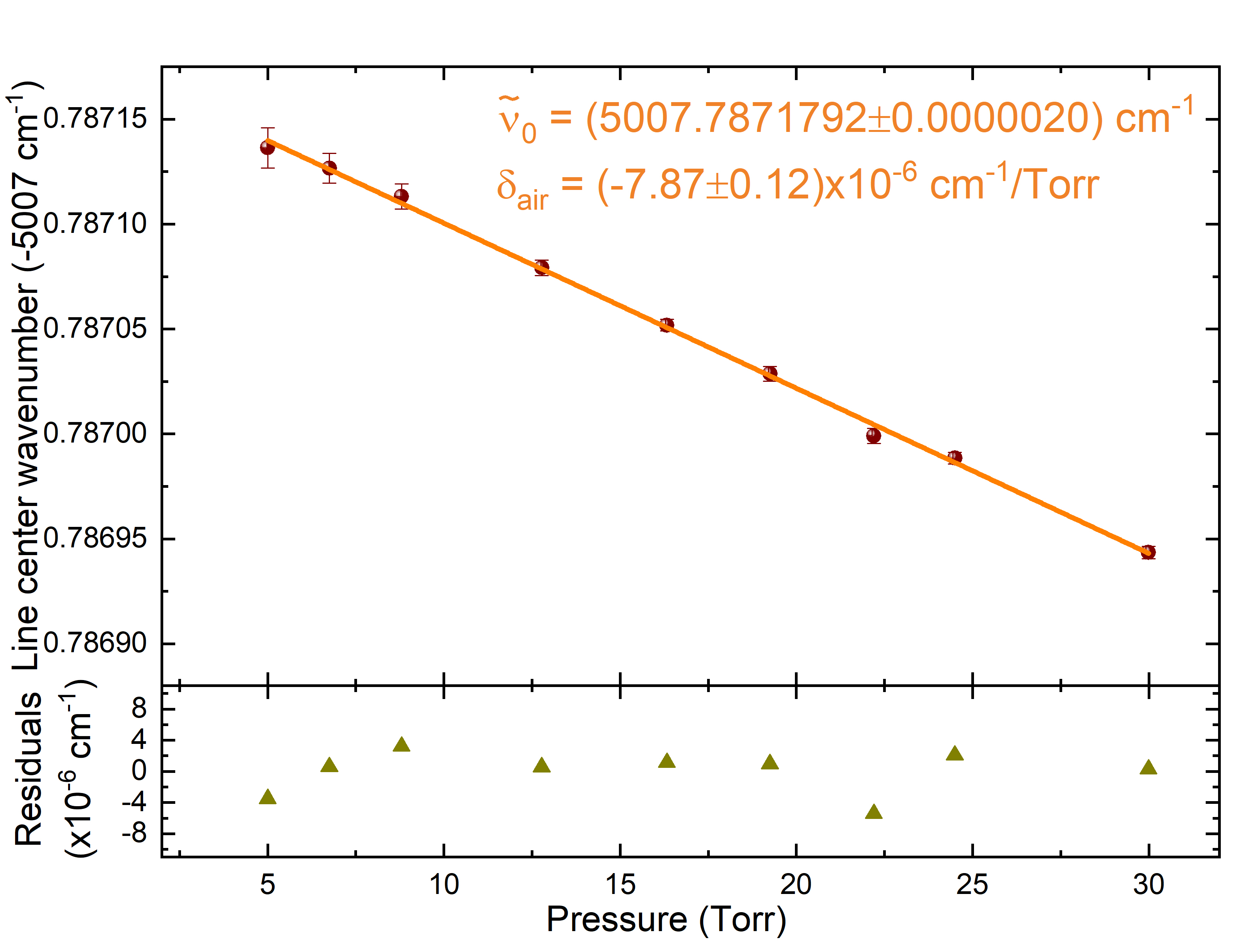}
    \caption{Upper panel: Pressure dependence of the line center frequency. The intercept and the slope of the best-fit line provide the zero-pressure line center frequency and the air-induced pressure shift coefficient, respectively. For clarity, on the vertical axis, the extrapolated zero-pressure ($\tilde{\nu}_0$) has been subtracted from the center frequency values retrieved from the fit. Lower panel: residuals from the linear fit, being their rms value 2.7$\times$10$^{-6}$ cm$^{-1}$.}
    \label{fig4b}
    \end{subfigure}
    \caption{Dependence on the gas pressure of the Lorentzian component and line center frequency of the investigated CO$_2$ transition.}
    \label{fig4}
\end{figure}

If for the investigated transition, we plot the
measured Lorentzian width $\Gamma_0$ versus the pressure and we fit the resulting points to a
weighted straight line (see Fig. \ref{fig4a}), we can determine the air-broadening coefficient, $\gamma_{air}$, namely the pressure-normalized $\Gamma_0$. In the linear best-fit procedure, we took into account the uncertainty on the collisional widths and pressures. In particular, the former represents the statistical contribution that arises from the spectra analysis, the latter comes from the type B uncertainty related to the accuracy of the pressure reading. We draw the reader’s attention to the nearly perfect agreement between the data points and the linear fit, characterized by a coefficient of determination (R$^2$) greater than 0.9998. In the case of Fig. \ref{fig4a}, $\gamma_{air}$ is equal to (8.64$\pm$0.04)$\times$10$^{-5}$ cm$^{-1}$/Torr, which is in excellent agreement, within the uncertainty, with the data provided by the HITRAN database \cite{HITRAN} and the CDSD-2024 databank \cite{CDSD-2024}, namely 8.59 and 8.60$\times$10$^{-5}$ cm$^{-1}$/Torr, respectively\footnote{The units of those parameters that are expressed in cm$^{-1}$/Torr can be converted in SI units, namely in MHz/Pa, using the identity 1 cm$^{-1}$/Torr = 2.2486$\times$10$^{2}$ MHz/Pa}. Most importantly, our experimental determination of $\gamma_{air}$ is fully consistent with the data recently obtained by Mondelain and co-workers in Ref. \cite{MONDELAIN2025}.

Similarly, as shown in Fig. \ref{fig4b}, a weighted linear fit of the center frequency as a function of the total gas pressure can be performed in order to extrapolate its zero-pressure value. In particular, the line center frequency was found to be (5007.7871792$\pm$0.0000020) cm$^{-1}$, the fractional statistical uncertainty being 4$\times$10$^{-9}$. It is worth noting that this value differs from the one listed in HITRAN of about 3 MHz. In spite of that, we can state that the two values are fully consistent, the uncertainty of the HITRAN data being between 3 and 30 MHz. The linear fit also provides the air-induced pressure shifting coefficient, $\delta_{air}$=(-7.87$\pm$0.12)$\times$10$^{-6}$ cm$^{-1}$/Torr, which is about 12\% lower than the corresponding HITRAN value. Nevertheless, the agreement may be considered satisfactory since the HITRAN value is the result of calculations based upon a simple empirical model proposed for the pressure-induced shifts of air-broadened CO$_2$ absorption lines \cite{HARTMANN2009}. Moreover, as for the comparison with the CDSD-2024 databank \cite{CDSD-2024}, we can state that our experimental value of $\tilde{\nu}_0$ is only $\sim$300 kHz higher, while $\delta_{air}$ maintains a difference that is similar to the one with the HITRAN database, namely, $\sim$12\%. It is worth noting that the last release of the ExoMol line list of the twelve isotopologues of CO$_2$ \cite{ExoMol}, for the R(50) transition, reports a line center frequency of 5007.78717 cm$^{-1}$, thus showing a negative shift of only 280 kHz compared to our experimental determination. 

Another interesting result is represented by the retrieved pressure-normalized $\nu_{opt}^r$ value that is in full agreement with those reported by Mondelain \textit{at al.} for transitions belonging to the same ro-vibrational band \cite{MONDELAIN2023}.

\begin{figure}[t]
\centering
\includegraphics[width=0.8\textwidth]{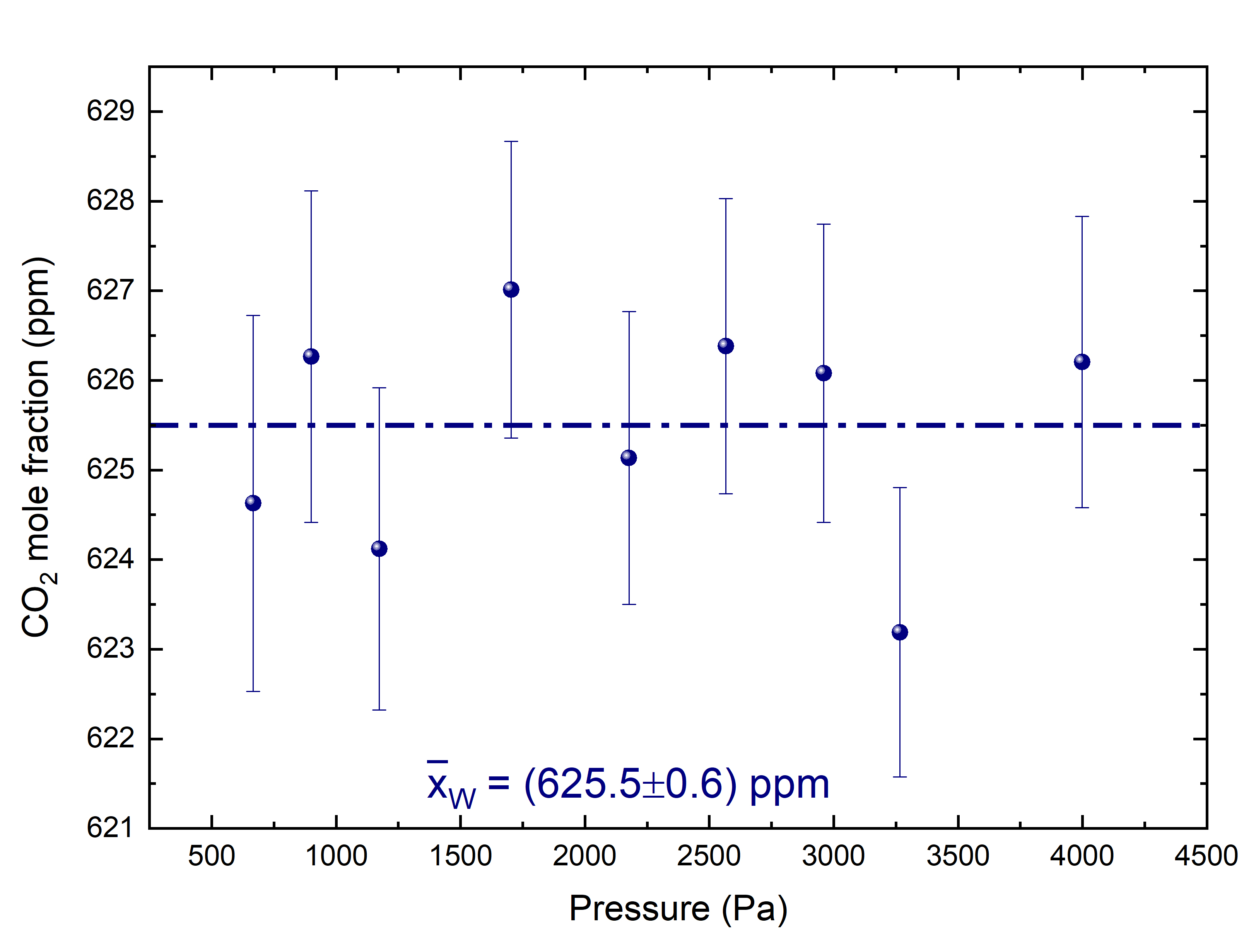}
\caption{CO$_2$ mole fraction determinations over the pressure range investigated in this work. The dashed-dotted line represents the $x_{CO_2}$ mean value.}
\label{fig5}
\end{figure}

It is interesting to note that the residuals of Figs. \ref{fig4a} and \ref{fig4b} do not show any particular structure, being normally distributed around a zero mean value. This is a further indication of the goodness of the linear fit. Albeit physical constrains have not been imposed, as usually done when a Multi Spectrum Fitting (MSF) procedure is employed \cite{Amodio2014, BENNER1995, Adkins2022-2}, we were able to capture the linear behavior of the Lorentzian width and line center frequency, despite the rather complicated line shape function that has been adopted. As for the pressure broadening and shifting coefficients and the zero-pressure line center frequency values, we compared the results obtained from the data of Figs. \ref{fig4a} and \ref{fig4b} with those retrieved adopting a MSF procedure. In this latter case, following the procedure deeply described in Ref. \cite{Amodio2014}, physical constraints were implemented for the collisional width, line center frequency, and real part of the Dicke narrowing parameter, while the Doppler width and the a$_W$ parameter were kept fixed. In more details, using the MSF procedure we found $\gamma_{air}$=(8.577$\pm$0.010)$\times$10$^{-5}$ cm$^{-1}$/Torr, $\delta_{air}$=(-7.74$\pm$0.12)$\times$10$^{-6}$ cm$^{-1}$/Torr, and $\tilde{\nu}_0$=(5007.7871794$\pm$0.0000020) cm$^{-1}$. Within their 2-$\sigma$ statistical uncertainty, these values are fully consistent with those derived in Figs. \ref{fig4a} and \ref{fig4b}.

As a further outcome of the line-fitting procedure, the CO$_2$ mole fraction, $x_{CO_2}$, could be determined, exploiting the fact that the transition
strength of the selected line, at the reference temperature of $T_{ref}$= 296 K, is known. Specifically, $S(T_{ref})$ is equal to 5.405$\times$10$^{-23}$ cm/molecule, as reported in the HITRAN database \cite{HITRAN}, which is also the one provided by the group of J. Tennyson at University College of London \cite{ZAK2016}. If the integrated absorption coefficient, $\alpha_{TOT}$, is normalized to the reference temperature using the lower-state energy level of the transition ($E''$=994.19130 cm$^{-1}$) and the calculated total partition functions, $Q(T)$ and $Q(T_{ref})$  \cite{GAMACHE2017}, the following equation holds:
\begin{equation}
\label{molefrac}
    x_{CO_2}=\frac{\alpha_{TOT}k_BT}{S(T_{ref})p},
\end{equation}
$p$ being the total gas pressure. For each spectrum of Fig. \ref{fig3}, the corresponding $x_{CO_2}$ value was retrieved. The outcomes of this analysis are summarized in Fig. \ref{fig5}. Here, the mean value was found to be (625.5$\pm$0.6) part-per-million (ppm). The latter was calculated as the weighted mean of the data, the error being the statistical uncertainty of the mean. It should be noted that, for each of the points of Fig. \ref{fig5}, the error bar includes the contributions arising from pressure and temperature reading, partition function, and statistical uncertainty on $\alpha_{TOT}$ as retrieved from the nonlinear fit. We draw the reader’s attention on the fact that the mean value of the mole fraction, which is shown in Fig. \ref{fig5}, is greater than the expected one, namely, the current atmospheric CO$_2$ concentration ($\approx$420 ppm). 
This is not surprising in a closed environment such as our laboratory, due to the presence of two or three people and limited air circulation. Moreover, it is worth noting that, if the CO$_2$ mole fractions are determined from the MSF procedure, the mean $x_{CO_2}$ value is fully consistent with the one formerly provided.

The limit of detection of our spectrometer, in terms of minimum detectable CO$_2$ mole fraction, $x^{min}_{{CO_2}}$, can be easily inferred, taking into account the SNR of the experimental spectra. In particular, under our experimental conditions, considering that the residuals' rms of Fig.\ref{fig3} is $\sim$ 1$\times$10$^{-9}$ cm$^{-1}$, a $x^{min}_{{CO_2}}$ of 0.15 ppm can be determined. This value, which obviously depends on the intensity of the probed transition, can be reduced by a factor of $\approx$40 by selecting one of the strongest components of the same band. In fact, CO$_2$ lines at 2-$\mu$m wavelength may reach an  intensity as high as 2$\times$10$^{-21}$ cm/molecule.

\section{Conclusions}
\label{concl}

In conclusion, we reported on spectroscopic parameters determinations of a carbon dioxide transition at 2 $\mu$m, using a new approach of comb-locked cavity ring-down spectroscopy. The proposed method is based on the use of an external-cavity diode laser that is offset-frequency locked to the signal output of a singly resonant optical parametric oscillator. This latter acts as a reference laser, being locked to a self-referenced optical frequency comb, which in turn is stabilized against a GPS-disciplined Rb-clock.

Absorption spectra, relative to the R(50) transition of the CO$_2$ 20012-00001 band at $\sim$5000 cm$^{-1}$, were observed with a high spectral fidelity, the latter being ensured by the absolute nature and high stability of the frequency axis. As a result of a careful spectral analysis, based on the modified Hartmann-Tran profile, we were able to determine the zero-pressure line center frequency of the transition with a relative uncertainty of 4$\times$10$^{-9}$. Other spectroscopic parameters, such as the air broadening and shifting coefficients, have been determined with a relative uncertainty  of 0.5 and 1.5\%, respectively. Furthermore, we demonstrate the measurement of the CO$_2$ mole fraction in ambient air gas samples with a sub-ppm uncertainty (namely, 0.09\% in relative terms).

The proposed methodology represents the basis for more advanced and intensive studies devoted to the precise and accurate determination of CO$_2$ spectroscopic parameters in the spectral window around 2 $\mu$m. In fact, with the aim of improving the results obtained by our group in 2017 \cite{Odintsova2017}, we are planning measurement campaigns on gravimetrically prepared gas
mixtures to further test \textit{ab-initio} quantum chemistry calculations of the CO$_2$ line intensities.

Similarly, the extension to SI-traceable $^{13}$C/$^{12}$C
isotope amount-ratio determinations in atmospheric CO$_2$ will be performed, following the seminal idea proposed in \cite{Castrillo2012} and the recent results of Ref. \cite{Srivastava2025}. Furthermore, the method can be applied to other molecular species of atmospheric interest, thanks to the wide tunability range of both the ECDL and OPO signal.

\section*{Funding}
This work was supported by the EURAMET project PriSpecTemp [grant number 22IEM03]. This project received funding within the Metrology Partnership Program co-financed
by the Participating States and the European Union’s Horizon 2020 research and motivation program.

\section*{CRediT authorship contribution statement}
\textbf{M.A. Khan}: Investigation, Data curation. \textbf{V. D'Agostino}: Investigation, Data curation. \textbf{S. Gravina}: Validation \& Investigation. \textbf{L. Gianfrani}: Conceptualization, Writing - original draft, review \& editing, Supervision. \textbf{A. Castrillo}: Conceptualization, Formal analysis, Writing - original draft, Writing - review \& editing, Funding acquisition.

\section*{Acknowledgment} 
This paper is written in memory of our friend, Misha Tretyakov, whose sudden departure left us with a deep sense of sadness. LG is grateful to him for fruitful interactions and, most importantly, selfless friendship over more than two decades. 

SG acknowledges the Italian Ministry for University and Research for providing a Research Fellowship through the ETIC Project within the PNRR program. 

\section*{Disclosures} 
The authors declare no conflicts of interest.

\section*{Data availability}
Data underlying the results presented in this paper are not publicly available at this time but may be obtained from the authors upon reasonable request.

\bibliographystyle{elsarticle-num} 

\end{document}